\documentstyle[12pt]{article}
\begin{document}
\begin{titlepage}

\hfill{May 1997}

\hfill{UM-P-97/31}

\hfill{RCHEP-97/05}

\vskip 0.9 cm

\centerline{{\large \bf 
How neutrino oscillations can induce an effective neutrino}}
\centerline{{\large \bf 
number of less than 3 during Big Bang Nucleosynthesis.  }}
\vskip 1.3 cm
\centerline{R. Foot and R. R. Volkas}

\vskip 1.0 cm
\noindent
\centerline{{\it School of Physics}}
\centerline{{\it Research Centre for High Energy Physics}}
\centerline{{\it The University of Melbourne}}
\centerline{{\it Parkville 3052 Australia }}

\vskip 2.0cm

\centerline{Abstract}
\vskip 0.7cm
\noindent
Ordinary-sterile neutrino oscillations can generate
significant neutrino asymmetry in the early Universe.
In this paper we extend this work by computing the
evolution of neutrino asymmetries and light element
abundances during the Big Bang Nucleosynthesis (BBN) epoch. 
We show that a significant electron-neutrino asymmetry can be
generated in a way that is approximately independent of the
oscillation parameters $\delta m^2$ and $\sin^2 2\theta$ for
a range of parameters in an interesting class of models. The 
numerical value of the
asymmetry leads to the {\it prediction} that the effective
number of neutrino flavours during BBN is either about 2.5
or 3.4 depending on the sign of the asymmetry.
Interestingly, one class of primordial deuterium abundance 
data favours an effective number of neutrino flavours during 
the epoch of BBN of less than 3.

\end{titlepage}

\vskip 0.8cm
\noindent
{\bf I. Introduction}
\vskip 0.5cm
The possible existence of sterile neutrinos can be motivated
by the solar neutrino, atmospheric neutrino 
and LSND experiments\cite{rev2}.
There are also interesting theoretical motivations for
the existence of light sterile neutrinos.
For example, if nature respects an exact unbroken parity
symmetry, then three necessarily light mirror neutrinos
must exist\cite{fv2}. In view of this, it is interesting
to study the implications of ordinary - sterile neutrino oscillations
for both particle physics and cosmology. 
The effects of ordinary - sterile neutrino oscillations
in the early Universe are actually quite remarkable.
It turns out that for a wide range of parameters, ordinary - 
sterile neutrino oscillations generate a large neutrino 
asymmetry \cite{ftv} (see also \cite{shi}). (A large neutrino 
asymmetry implies that the universe has a net nonzero lepton number
given that the electron asymmetry is necessarily small due to
the charge neutrality requirement.) One important
implication of this result is that the bounds on ordinary - sterile
oscillation parameters that can be derived mainly from energy density
considerations during Big Bang Nucleosynthesis (BBN) 
are severely affected  (see Ref.\cite{fv} for a detailed analysis).
However, electron lepton number can also affect BBN
directly through the modification of nuclear reaction rates.
It is this issue that we will study in this paper.

In a previous paper\cite{fv}, we showed that for a wide range of 
parameters, the evolution of lepton number can be approximately
described by a relatively simple first order integro-differential
equation. We called the approximation used there 
the ``static approximation''
because it holds provided that the system is sufficiently smooth. 
The static approximation is valid in the region where
the evolution of lepton number is dominated by collisions.
In particular, for the temperature at which 
lepton number is initially
produced, this approximation is generally valid for $|\delta m^2|
\stackrel{>}{\sim} 10^{-2}\ eV^2$\cite{fv}.  However, it 
is not expected to be valid for 
temperatures much less than the temperature at which lepton number
is initially generated.
This is because the static approximation discussed in Ref.\cite{fv} 
does not incorporate the MSW effect\cite{msw}, which
is in fact the dominant
process affecting the evolution of lepton number at low
temperatures.  For the application considered in Ref.\cite{fv}, 
the evolution of lepton number at low temperatures was not required. 
However, for the application of the present paper the accurate 
evolution 
of lepton number to temperatures $T \sim 1 \ MeV$ is necessary
in order to study its precise effect on BBN reaction rates.

The outline of this paper is as follows.  In section II we set 
the scene with a brief review of the effects of neutrino asymmetry 
on BBN.  In section III we develop a simple formalism describing
the evolution of lepton number at low temperatures where
the MSW effect is important. This work can also be
viewed as an extension of our previous study\cite{fv},
where the evolution of lepton number at higher temperature
was studied in detail.  In this section we also examine
the implications for BBN of direct electron asymmetry generation
by  $\nu_e - \nu_s$ oscillations.
In section IV we examine a more interesting scenario
where electron asymmetry is generated indirectly.
In section V, we provide a check of our simple formalism (of section
III) by numerically solving the exact quantum kinetic equations.
Finally in section VI we conclude.

\vskip 0.5cm
\noindent
{\bf II Electron neutrino asymmetry and BBN}
\vskip 0.5cm
Standard BBN can give a prediction for
the effective number of neutrino flavours, $N_{\nu}^{eff}$, present
during nucleosynthesis. This prediction depends on 
the baryon to photon number-density ratio, $\eta$, and the primordial 
helium mass fraction, $Y_P$.  A precise determination of the 
primordial deuterium abundance will provide a quite sensitive measurement 
of $\eta$.  Once $\eta$ is known, the effective number of neutrinos 
present during nucleosynthesis depends only on $Y_P$.  At present 
there are two conflicting deuterium observations in different 
high-redshift low-metallicity quasistellar object absorbers. There is 
the high deuterium result of Ref.\cite{rh} which suggests that 
$D/H = (1.9 \pm 0.4) \times 10^{-4}$.
On the other hand, in Ref.\cite{bt} the low deuterium result 
of $D/H = [2.3 \pm 0.3(stat) \pm 0.3 (syst)]  \times 10^{-5}$
is obtained.  The implications of these results for the prediction
of $N_{\nu}^{eff}$ have been discussed in
a number of recent papers\cite{papers}. Depending on which 
of these two values of the deuterium 
abundance is assumed, different predictions for $\eta$ are obtained.
The high deuterium result leads to $\eta \sim 2 \times 10^{-10}$,
while the low deuterium result leads to $\eta \sim 7 \times 
10^{-10}$\cite{papers}.  Each of these predictions for 
$\eta$, together with the inferred primordial abundance of 
$^4He$, allows a prediction for $N^{eff}_{\nu}$
to be made\cite{papers}. According to Ref.\cite{hata2} for 
example, the high deuterium case leads to
\begin{equation}
N_{\nu}^{eff} = 2.9 \pm 0.3, 
\end{equation}
while the low deuterium case leads to
\begin{equation}
N_{\nu}^{eff} = 1.9 \pm 0.3,
\label{ss}
\end{equation}
where the errors are at $68\% $ C.L. 
The minimal standard model of course predicts that $N_{\nu}^{eff}
= 3$. Thus, if the low deuterium result were correct, then
new physics would presumably be required\cite{h95}.  Of course, 
estimating the primordial element abundances is difficult and it is 
possible that the primordial helium abundance has been 
underestimated (in otherwords, even if the low
deuterium measurement is correct $N^{eff}_{\nu} = 3$ is
not inconsistent).  
Fortunately the situation is continually improving as more observations
and analyses are done.
In the interim it is useful to identify and study the types
of particle physics that can lead to $N_{\nu}^{eff} < 3$.

One possibility is that the electron lepton number is
large enough to significantly affect BBN 
(i.e. $L_{\nu_e} \stackrel{>}{\sim} 0.01$)\cite{1}.
The relationship between an electron neutrino asymmetry and 
the effective
number of neutrino species arises as follows. A nonzero 
electron neutrino asymmetry modifies the nucleon reaction rates 
($n + \nu_e \leftrightarrow p + e^-$, $n + e^+ \leftrightarrow
p + \bar \nu_e$) which keep the neutrons and protons in thermal 
equilibrium down to temperatures of about $0.7\ MeV$. 
A modification of these rates affects the ratio
of neutrons/protons and hence  changes the prediction 
of $Y_P$\cite{1}. A change of $Y_P$ can be equivalently 
expressed as a change in the effective number
of neutrino species, $\delta N_{\nu}^{eff}$, present
during nucleosynthesis. These quantities are related by the equation
(see e.g. Ref.\cite{walker})
\begin{equation}
\delta Y_P \simeq 0.012\delta N_{\nu}^{eff}.
\label{2}
\end{equation}
The effect of the electron neutrino asymmetry on the primordial
helium abundance is most important in the temperature region  
\begin{equation}
0.4 \ MeV \stackrel{<}{\sim} T \stackrel{<}{\sim} 1.5 \ MeV
\label{range2}
\end{equation}
where the reactions $n + \nu_e \leftrightarrow p + e^-$
and $n + e^+ \leftrightarrow p + \bar \nu_e$ fix the neutron/proton 
ratio. For temperatures less than about $0.4 \ MeV$, these
reaction rates become so slow that the 
dominant process affecting the neutron/proton ratio is neutron decay.
Note that the Helium mass fraction $Y_P$ satisfies the differential
equation\cite{rev},
\begin{equation}
{dY_P \over dt} = - \lambda (n \to p)Y_P + \lambda(p \to n)
(2 - Y_P),
\label{yy}
\end{equation}
where $\lambda (n \to p)$ [$\lambda (p \to n)$] is 
the rate at which neutrons are converted into protons [protons are
converted to neutrons]. For temperatures in the range of 
Eq.(\ref{range2}),
$\lambda (n \to p) \simeq \lambda(n + \nu_e \to
p + e^-) + \lambda (n + e^+ \to p + \bar \nu_e)$ and
$\lambda (p \to n) \simeq \lambda(p + \bar \nu_e \to
n + e^+) + \lambda (p + e^- \to n + \nu_e)$.
The reaction rates (per nucleon) are obtained by 
integrating the square of
the matrix element weighted by the available phase-space.
For example, the rate for the process $n + \nu_e \to p + e^-$
is given by
\begin{equation}
\lambda (n + \nu_e \to p + e^-) = 
\int f_{\nu}(E_{\nu})[1 - f_e(E_e)]
|{\cal M}|^2_{n\nu_e \to p e}(2\pi)^{-5}
\delta^4 (n + \nu - p - e)
{d^3 p_{\nu} \over 2E_{\nu}}
{d^3 p_{e} \over 2E_e}
{d^3 p_{p} \over 2E_p}
\end{equation}
where $f_i (E_i)$ is the Fermi-Dirac distribution 
$f_i(E_i) \equiv [exp(E_i/T) + 1]^{-1}$.
These reaction rates are modified in the presence of significant 
electron neutrino asymmetry.
If the neutrino asymmetry is produced at temperatures above
about $1.5$ MeV and is constant over the temperature range of
Eq.(\ref{range2}) then we only need to add in the appropriate
chemical potentials $\mu_{\nu}$ and $\mu_{\bar \nu}$ 
to the distributions $f_{\nu}$ and $f_{\bar \nu}$. 
\vskip 0.5cm
\noindent
{\bf III Neutrino oscillation generated neutrino asymmetry}
\vskip 0.5cm

We now discuss the effects of neutrino oscillations, assuming that
a sterile neutrino exists.
Our convention for the neutrino oscillation parameters, 
$\delta m_{\alpha s}^2$
and $\sin^2 2\theta_0$, is as follows.
For $\nu_{\alpha} -\nu_{s}$ oscillations (with $\alpha = e, \mu, \tau$),
the weak eigenstates $\nu_{\alpha}$ and $\nu_s$ are linear
combinations of mass eigenstates $\nu_a$ and $\nu_b$,
\begin{equation}
\nu_{\alpha} = \cos\theta_0 \nu_a + \sin\theta_0 \nu_b, 
\ \nu_s = -\sin\theta_0 \nu_a + \cos\theta_0 \nu_b, 
\end{equation}
where the vacuum mixing angle $\theta_0$ is 
defined so that $\cos 2\theta_0 \ge 0$.
Further, we define the oscillation parameter
$\delta m^2_{\alpha s}$ by $\delta m^2_{\alpha s} 
\equiv m_b^2 - m_a^2$.
Also, the term ``neutrino'' will sometimes
include anti-neutrino as well. We hope that the correct
meaning will be clear from context.

Ordinary - sterile neutrino oscillations can create
significant lepton number provided that $\delta m^2_{\alpha s} < 0$ and
$|\delta m_{\alpha s}^2| \stackrel{>}{\sim} 10^{-4} \ eV^2$.
For full details, see Refs.\cite{ftv, shi, fv}.
In the following we consider $\nu_{\alpha} - \nu_s$ oscillations
in isolation. It is important to note that this is not in
general valid because the effective potential (see below) 
depends on all of the lepton asymmetries. However, it 
is approximately valid for the ordinary - sterile neutrino
oscillations which have the largest $|\delta m^2|$\cite{fv}.

The effective potential describing the coherent
forward scattering of neutrinos of momentum $p \equiv |\vec{p}|
\simeq E$ with the background is\cite{nr,msw}
\begin{equation}
V_{\alpha} \equiv V_{\alpha}(T,p,L^{(\alpha)}) 
= {\delta m^2_{\alpha s} \over 2p}(-a + b),
\end{equation}
where the dimensionless functions
$a$ and $b$ are given by
\begin{equation}
a \equiv a(T,p,L^{(\alpha)}) = 
{{-4 \zeta(3) \sqrt{2} G_F T^3 L^{(\alpha)} p} 
\over {\pi^2 \delta m_{\alpha s}^2}},\ \ 
b \equiv b(T,p) = {{-4 \zeta(3) \sqrt{2} G_F T^4 A_{\alpha} p^2}
\over {\pi^2\delta m_{\alpha s}^2 M_W^2}},
\label{ab}
\end{equation}
and $\zeta (3) \simeq 1.202$ is the Riemann zeta function
of 3, $G_F$ is the Fermi constant, $M_W$ is the W-boson mass, 
$A_e \simeq 17$ and $A_{\mu,\tau} \simeq 4.9$ \cite{notation}.
The quantity $L^{(\alpha)}$ is given by
\begin{equation}
L^{(\alpha)} = L_{\nu_{\alpha}} + L_{\nu_e} + L_{\nu_{\mu}} + 
L_{\nu_{\tau}} + \eta,
\label{hhi}
\end{equation}
where $L_{\nu_{\alpha}} \equiv (n_{\nu_{\alpha}} - n_{\bar
\nu_{\alpha}})/n_{\gamma}$ with $n_i$ being the 
number density of species $i$. 
In kinetic equilibrium $n_i$ and hence $L^{(\alpha)}$
is in general a function of the independent variables $\mu_i$ 
(the chemical potential) and $T$. For the situation we will be
considering, the asymmetry $L^{(\alpha)}$ quickly becomes
independent of its initial value (see Ref.\cite{fv} for
a complete discussion). This effectively means that
$\mu_i$ is not an independent variable but rather it
becomes a function of $T$. The asymmetry $L^{(\alpha)}$
is thus essentially
a function of $T$ only, and $a$ and $b$ are
functions of $p$ and $T$ only. The quantity $\eta \simeq L_N/2$ 
is a small term ($\sim 10^{-10}$) which 
arises from the asymmetries of baryons and electrons. 
The matter mixing angles are expressed in terms of
the quantities $a$ and $b$ through\cite{msw}
\begin{eqnarray}
\sin^2 2\theta_m & \equiv & 
\sin^2 2\theta_m (T,p,L^{(\alpha)})
= {\sin^2 2\theta_0 \over
{\sin^2 2\theta_0 + (b - a - \cos2\theta_0)^2}},\ \nonumber\\
\sin^2 2\bar \theta_m & \equiv & 
\sin^2 2\bar \theta_m(T,p,L^{(\alpha)})
= {\sin^2 2\theta_0 \over
{\sin^2 2\theta_0 + (b + a - \cos2\theta_0)^2}}.
\label{dons}
\end{eqnarray}
Note that the MSW resonance occurs for neutrinos of
momentum $p$ when $\theta_m = \pi/4$
and for antineutrinos of momentum $p$
when $\bar \theta_m = \pi/4$, which from
Eq.(\ref{dons}) implies that $b - a = \cos2\theta_0$ and
$b + a = \cos2\theta_0$, respectively.

If $\sin^2 2\theta_0 \ll 1$, then it can be shown that
oscillations with $b < 1$ create lepton number
while the oscillations with $b > 1$ destroy lepton number\cite{fv,ftv}.
Since $\langle b \rangle \sim T^6$, 
it follows that at some point
the lepton number creating oscillations dominate over
the lepton number destroying oscillations (where the brackets
$\langle\cdots\rangle$ denote the thermal momentum average).
We will call this temperature $T_c$. It is given
roughly by the temperature where $\langle b \rangle
= \cos2\theta_0 \simeq 1$, that is
\begin{equation}
T_c \simeq 13(16)\left({|\delta m_{\alpha s}^2| \over eV^2}
\right)^{1\over 6}\ MeV,
\label{Tc}
\end{equation}
for $\nu_e - \nu_s$ ($\nu_{\mu,\tau} -\nu_s$) oscillations.

It is important to note that there are 
two distinct contributions to
the rate of change of lepton number. One contribution is from
the oscillations between collisions. The other is from
the collisions themselves which deplete neutrinos and
anti-neutrinos at different rates in a CP asymmetric background.
It turns out that for $T \stackrel{>}{\sim} T_c$, lepton number
evolution is dominated by collisions for the small
vacuum mixing angle case we are considering assuming 
that $|\delta m^2_{\alpha s}| \stackrel{>}{\sim} 10^{-4}
\ eV^2$\cite{ftv, shi, fv, bul}. Oscillations 
between collisions, and in particular MSW transitions, can be 
ignored for $T \stackrel{>}{\sim} T_c$ because the interactions are 
so rapid that the mean distance
between collisions $L_{int}$ is much smaller than the 
matter-oscillation length $L_{osc}^m$ (and consequently
the neutrino cannot evolve coherently through the
resonance). To see this note that the amplitude of the
oscillations at the MSW resonance is given roughly by 
$\sin^2 L_{int}/2L_{osc}^m$, where $L_{int} \sim 
1/(G_F^2p_{res}T^4)$ is the interaction length at the resonance
and $L_{osc}^m \sim 2p_{res}/(\sin2\theta_0 
|\delta m_{\alpha s}^2|)$ 
is the oscillation length at the resonance\cite{fv}. 
So, at the resonance,
\begin{equation}
{L_{int}\over L_{osc}^m} \sim
{\sin2\theta_0 \over {G_F^2 p_{res} T^4}}
{|\delta m_{\alpha s}^2| \over 2p_{res}}
\simeq 10^2 \sin 2\theta_0  \left[{T_c \over T}\right]^6, 
\end{equation}
where we have used the approximation 
$p_{res} \simeq \langle p \rangle \simeq 3.15T$ for
the resonance momentum. Thus
$\sin^2 L_{int}/2L^m_{osc} \ll 1$ for $T \stackrel{>}{\sim}
T_c$, provided that $\sin^2 2\theta_0 \stackrel{<}{\sim} 10^{-4}$,
and so the MSW transitions are heavily suppressed. 
However, $L_{int}/L_{osc}^m \sim 1/T^6$  rapidly 
increases as $T$ becomes lower. Also, for $T \stackrel{<}{\sim}
T_c$ it turns out that $p_{res}/T \stackrel{<}{\sim} 0.8$ (see later). 
Taking these factors into account, 
$L_{int}/L^m_{osc} \stackrel{>}{\sim} 1$ for $T \stackrel{<}{\sim}
T_c/2$, provided that
$\sin^2 2\theta_0 \stackrel{>}{\sim} 3\times 10^{-10}$.
In this case the MSW effect will not be suppressed if the 
oscillations are adiabatic.
Furthermore the oscillations are generally
adiabatic
for the parameter space of interest in this paper
(which turns out to be $|\delta m^2_{\alpha s}|
\stackrel{>}{\sim} 10^{-1} \ eV^2$).

The key task in this paper is to analyse the evolution of lepton
number for $T \stackrel{<}{\sim} T_c/2$ when MSW transitions become
important. 
(The effect of MSW transitions was noted in Ref\cite{fv}, but 
since the evolution of
lepton number to low temperatures was not required for the 
application considered there, we did not study the
effect of MSW transitions in any detail, except to note that
they keep $L_{\nu_{\alpha}}$ growing like $1/T^4$ for
low temperatures). Now, through the quantum kinetic 
equations \cite{bm}, the evolution of lepton number can be calculated
in a close to exact manner, including the
effects of both collisions and oscillations between collisions. 
However, these complicated coupled
equations have two notable drawbacks. First, they do not furnish 
as much physical insight as one might wish. Second, they are
impractically complicated when one wishes to consider a 
system of more than two neutrino flavours. Since the physics
of MSW transitions is the essence of how lepton number evolves
during the BBN epoch, and since we will later need to consider
a system of four neutrino flavours, we now pursue a very
useful approximate approach instead of employing the
full quantum kinetic equations. We will along the way
check the veracity of
our approximate approach by comparing results with those
obtained from the quantum kinetic equations in the two flavour
case (see section V). 
This will give us confidence in the use of our 
appproximate formalism in the four-flavour case considered
later in this paper.

For definiteness we will assume that the
lepton number created at the temperature $T =T_c$ is positive in sign
\cite{fn}. In this case note that $a, b > 0$ given also that
$\delta m_{\alpha s}^2 < 0$.  
The momentum of the anti-neutrino oscillation 
resonance (obtained from the condition 
$b + a = \cos2\theta_0 \simeq 1$) typically moves to quite low values, 
$p_{res}/T \stackrel{<}{\sim} 0.8$ (for $T \stackrel{<}{\sim} T_c$).
In contrast, the neutrino oscillation resonance momentum obtained
from $- a + b \simeq 1$
moves to a very high value, $p_{res}/T \gg 1$ (see Figure 2 in
section V for an illustration of this). 
As $b \simeq \langle b \rangle \sim T^6$
becomes smaller, the neutrino momentum resonance $p_{res}/T$ 
very quickly becomes so high that
its effects can be neglected, because the resonance occurs
in the tail of the neutrino momentum distribution.  
Thus, for $T < T_c$, we can to a
a good approximation ignore the neutrinos,
and simply study the effects of the MSW transitions on
the anti-neutrinos.  In this case all the anti-neutrinos
which pass through the resonance are converted into sterile neutrinos
and vice-versa (the MSW effect). The rate of change of lepton 
number is thus related to the number of anti-neutrinos minus the number 
of sterile anti-neutrinos which pass through the resonance.  
Note that this rate is independent of the precise value
of $\sin^2 2\theta_0$ provided that
$\sin^2 2\theta_0 \ll 1$. Under this assumption
\begin{equation}
{dL_{\nu_{\alpha}} \over dT} = 
- \left({N_{{\bar\nu}_{\alpha}} - 
N_{{\bar \nu}_s} \over n_{\gamma}}\right) 
T |{d \over dT}\left({p_{res} \over T}\right)|,
\label{cat}
\end{equation}
where $N_i$ describes the momentum
distribution of species $i$, 
so that $n_i = \int_{0}^{\infty} N_i dp$. 
In thermal equilibrium,
\begin{equation}
N_{\bar \nu_{\alpha}} = {1 \over 2\pi^2}{p^2 \over 
1 + exp\left({p+\mu_{\bar \mu}\over T}\right)}.
\end{equation}
In Eq.(\ref{cat}) the factor
$T d(p_{res}/T)/dT \simeq dp_{res}/dT - p/T = 
dp_{res}/dT - dp/dT$ is the rate
at which $p_{res}$ changes relative to the neutrino momentum 
(for neutrinos with momentum $p \sim p_{res}$). 
Note that $N_{\nu_s} \simeq 0$ if the number-density 
of sterile neutrinos is negligible. The functions $N_i$ in
Eq.(\ref{cat}) are evaluated at the resonance momentum, $p_{res}$,
obtained as a function of $L_{\nu_{\alpha}}(T)$ and $T$
from the resonance condition $a \simeq \cos2\theta_0 \simeq 1$,
\begin{equation}
{p_{res} \over T} \equiv {p_{res}[T,L_{\nu_\alpha}(T)] \over T}  
= {-\pi^2 \delta m_{\alpha s}^2 \over 
8\zeta(3)\sqrt{2}G_FT^4 L_{\nu_{\alpha}}},
\label{hird}
\end{equation} 
where we have considered the case 
$\eta, L_{\nu_{\beta}} \ll L_{\nu_{\alpha}}$ for $\beta \neq \alpha$.
Note that this expression is only valid for $T \stackrel{<}{\sim} 
T_c/2$ where the $b$-term can be neglected.  Using
\begin{equation}
{d(p_{res}/T)\over dT} = {\partial(p_{res}/T) \over \partial T} + 
{\partial (p_{res}/T) \over \partial L_{\nu_{\alpha}}}
{dL_{\nu_{\alpha}}\over dT} = 
- 4 {p_{res} \over T^2} - {p_{res} \over T L_{\nu_\alpha}}
{dL_{\nu_\alpha} \over dT},
\end{equation} 
Eq.(\ref{cat}) yields
\begin{equation}
{dL_{\nu_{\alpha}}\over dT} = {{-4Xp_{res}/T} \over T + Xp_{res}/
L_{\nu_{\alpha}}},
\label{cat2}
\end{equation}
where we have assumed $d(p_{res}/T)/dT < 0$.
The useful dimensionless quantity $X$ is given by
\begin{equation}
X \equiv X[T, p, \mu_{{\bar \nu}_\alpha}(T), N_{\bar \nu_s}(T)]
= {T \over n_{\gamma}}\left(N_{{\bar \nu}_{\alpha}}
- N_{{\bar \nu}_s}\right)
\end{equation}
and it is evaluated at $p = p_{res}$.

Equation \ref{cat2} is a non-linear equation in $L_{\nu_\alpha}$.
The righthand side of this equation depends on $L_{\nu_\alpha}$
through $p_{res}$ directly, through the dependence of $X$ on $p_{res}$
and through the number densities. In order to solve this equation, 
we need to write the
chemical potentials in terms of $L_{\nu_\alpha}$.
Now, for each temperature $T$, the neutrino asymmetry is
created at the neutrino momentum $p_{res}$. However,
for temperatures greater than about $1 \ MeV$\cite{fn3}
the effect of the weak interactions is to quickly thermalise the
neutrino momentum distributions. This means that the neutrino
asymmetry is approximately distributed throughout the neutrino momentum
spectrum via chemical potentials for
the neutrinos and anti-neutrinos.  In general,
\begin{equation}
L_{\nu_{\alpha}} = {1 \over 4\zeta (3)}\int^{\infty}_0
{x^2dx \over 1 + e^{x+\stackrel{\sim}{\mu}_{\nu}}}
- {1 \over 4\zeta (3)}\int^{\infty}_0
{x^2dx \over 1 + e^{x+\stackrel{\sim}{\mu}_{\bar \nu}}},
\label{essendon for premiers}
\end{equation}
where $\stackrel{\sim}{\mu}_{i} \equiv \mu_{i}/T$
and $i = \nu, \bar \nu$. Expanding Eq.(\ref{essendon for premiers}), 
we find that
\begin{equation}
L_{\nu_{\alpha}} \simeq - {1 \over 24\zeta (3) }
\left[ \pi^2 (\stackrel{\sim}{\mu}_{\nu} -
\stackrel{\sim}{\mu}_{\bar \nu}) - 6(\stackrel{\sim}{\mu}_{\nu}^2 -
\stackrel{\sim}{\mu}_{\bar \nu}^2)\ln 2 + 
(\stackrel{\sim}{\mu}_{\nu}^3 - \stackrel{\sim}{\mu}_{\bar \nu}^3)
\right],
\end{equation}
which is an exact equation for $\stackrel{\sim}{\mu}_{\nu} 
= - \stackrel{\sim}{\mu}_{\bar \nu}$,
otherwise it holds to a good approximation
provided that $\stackrel{\sim}{\mu}_{\nu,\bar \nu}$ 
$\stackrel{<}{\sim}$ $1$.
For $T \stackrel{>}{\sim} T^{\alpha}_{dec}$ 
(where $T^e_{dec} \simeq 3\ MeV$ and 
$T^{\mu, \tau}_{dec} \simeq 5\ MeV$ are the chemical decoupling
temperatures) $\mu_{\nu_{\alpha}}
\simeq -\mu_{\bar \nu_{\alpha}}$ because processes such as
$\nu_{\alpha} + \bar \nu_{\alpha} \leftrightarrow 
e^+ + e^-$ are rapid enough
to make $\stackrel{\sim}{\mu}_{\nu} + \stackrel{\sim}{\mu}_{\bar
\nu}$ $\simeq$ $\stackrel{\sim}{\mu}_{e^+} + \stackrel{\sim}{\mu}_{e^-}$
$\simeq$ $0$.  However, for $1\ MeV \stackrel{<}{\sim} T \stackrel{<}{\sim}
T^{\alpha}_{dec}$, weak interactions are rapid 
enough to approximately thermalise the neutrino momentum distributions, 
but not rapid enough to keep the neutrinos in chemical equilibrium.
In this case, the value of $\stackrel{\sim}{\mu}_{\nu}$
is approximately frozen at $T \simeq T^{\alpha}_{dec}$,
while the anti-neutrino chemical potential 
$\stackrel{\sim}{\mu}_{\bar \nu}$ continues increasing
until $T \simeq 1\ MeV$.

We also need to specify the initial condition in order to solve
Eq.(\ref{cat2}). To do so we need to know the value of $L_{\nu_\alpha}$
at some temperature $T_i < T_c$ at which MSW transitions
are already dominant.
This $L_{\nu_\alpha}$ value can be obtained 
by solving the exact quantum kinetic
equations, based on the density matrix, which incorporate both
collision and oscillation effects. Fortunately, it turns out
that the subsequent evolution of lepton number is reasonably
insensitive to what temperature $T_i$ is chosen as the initial
temperature for Eq.(\ref{cat2}), provided that $T_i$
is chosen during the epoch after $T_c$ 
for which $L_{\nu_\alpha} \ll 1$\cite{binh}.
When the asymmetry $L_{\nu_{\alpha}} \ll 1$, 
Eq.(\ref{cat2}) can be simplified to
$dL_{\nu_{\alpha}}/dT \simeq -4L_{\nu_{\alpha}}/T$
which means that $L_{\nu_{\alpha}}T^4$ is approximately
constant. This means that $p_{res}/T$ is also approximately
constant given that $L_{\nu_{\alpha}}$ is related to the
resonance momentum $p_{res}$ by Eq.(\ref{hird}).
As we will discuss in section V, a numerical solution 
of the quantum kinetic equations shows that
$p_{res}/T$ is generally in the range
\begin{equation}
0.2 \stackrel{<}{\sim} p_{res}/T \stackrel{<}{\sim} 0.8,
\label{range}
\end{equation}
for $T$ values around
$T_i = T_c/2$ when the oscillation parameters have been
chosen to lie in the parameter space of interest,
which turns out to be 
$|\delta m_{\alpha s}^2| \stackrel{>}{\sim}
10^{-1}\  eV^2$ and $\sin^2 2\theta_0 \stackrel{>}{\sim} 5\times
10^{-10}(eV^2/|\delta m_{\alpha s}^2|)^{1/6}$. 
(This lower bound for the mixing 
angle ensures that a suitably large asymmetry is created at
$T_c$ \cite{fv}. The result of Eq.(\ref{range}) can also be gleaned
from the static approximation based results of Ref.\cite{fv}.)
We will from now on use a value of about $T_c/2$ for $T_i$.

Before presenting the results of a numerical solution of 
Eq.(\ref{cat2}) for the final asymmetry $L_{\nu_\alpha}$, 
it is interesting to note that an approximately correct
answer is easily obtained from the following argument.
As $T$ falls below $T_c/2$, the asymmetry keeps increasing.
This eventually forces the rate of change of $L_{\nu_\alpha}$
to decrease substantially. Recall that $dL_{\nu_\alpha}/dT$
is proportional to the how quickly the resonance momentum
$p_{res}$ moves as per Eq.(\ref{cat}). When $L_{\nu_\alpha}$
is large, $p_{res}/T$ must move to
large values in order to create lepton number.
Eventually, $p_{res}/T \to \infty$, and all
of the anti-neutrinos which have passed through the resonance 
have been converted into sterile neutrinos.
Thus, assuming that the initial number of sterile neutrinos
is negligible and also neglecting the modification
of the distribution due to the chemical potential, we expect that
the final value of the lepton number, $L^f_{\nu_{\alpha}}$,
to be given roughly by
\begin{equation}
{L^f_{\nu_{\alpha}} \over h}
\simeq {1 \over 4\zeta(3)}\int^{\infty}_{p_{in}/T} 
{x^2 dx \over 1 + e^x} 
\simeq {3 \over 8},
\label{cat4}
\end{equation}
where $h = T_{\nu_{\alpha}}^3/T_{\gamma}^3$ (note that $h \simeq 1$ for
$T \stackrel{>}{\sim} m_e \simeq 0.5 \ MeV$) and $p_{in}/T$ is the
value of $p_{res}/T$ [and is in the range of Eq.(\ref{range})] at
$T \simeq T_c/2$.
It is interesting that the final asymmetry is approximately
independent of $p_{in}/T$ and hence also of $\delta m_{\alpha s}^2$.
This is because $p_{in}/T$ from Eq.(\ref{range}) is always small.

Actually, the final value of the lepton 
number is somewhat less than $3/8 = 0.375$ if it is created when 
$T \stackrel{>}{\sim} 1$ MeV.  This is because the number density of 
anti-neutrinos is continually reduced as the lepton number is thermally
distributed via the chemical potential.
Thus, $L_{\nu_{\alpha}}^f$ depends on
the temperature at which $L_{\nu_{\alpha}}$ becomes large 
($10^{-2}$ roughly), and thus on $|\delta m_{\alpha s}^2|$.  
Numerically solving Eq.(\ref{cat2}), assuming
that the initial number of sterile neutrinos is negligible, we 
found that the final value of the lepton number is\cite{fn4}
\begin{eqnarray}
L^f_{\nu_{\alpha}}/h \simeq 0.35\ for \ 
|\delta m_{\alpha s}^2|/eV^2 \stackrel{<}{\sim}
3, \nonumber \\
L^f_{\nu_{\alpha}}/h \simeq 0.23\ for \ 3 \stackrel{<}{\sim} 
|\delta m_{\alpha s}^2|/eV^2 \stackrel{<}{\sim}
3000,  \nonumber \\
L^f_{\nu_{\alpha}}/h \simeq 0.29\ for \ 
|\delta m_{\alpha s}^2|/eV^2 \stackrel{>}{\sim} 3000.
\end{eqnarray}
In numerically solving Eq.(\ref{cat2}) we start the
evolution at $T \simeq T_c/2$ with $p_{in}/T$ in the range of
Eq.(\ref{range}) and with a corresponding $L_{\nu_\alpha}$ obtained
through Eq.(\ref{hird}). We found that $L^f_{\nu_{\alpha}}$
is approximately independent of the initial value of 
$L_{\nu_{\alpha}}$ for $p_{in}/T$ in this range.

The temperature where the final neutrino asymmetry
is reached is approximately, 
\begin{equation}
T^f_{\nu} \simeq
0.5\left(|\delta m^2|/eV^2\right)^{1/4}\ MeV.
\end{equation}
This result can be obtained analytically by using the 
resonance relation Eq.(\ref{hird}) with $L_{\nu_{\alpha}}
\simeq L^f_{\nu_{\alpha}}$ and $p_{res}/T \sim 6$ (since
$L^f_{\nu_{\alpha}}$ is not reached until 
$p_{res}/T \gg 1$ and we take $p_{res}/T \sim 6$
for definiteness).

Equation \ref{cat2} is an approximation based on the neglect of
collisions and the assumption of complete MSW conversion.
By numerically integrating the exact quantum kinetic
equations\cite{bm}, we have checked that Eq.(\ref{cat2}) does indeed 
accurately describe the evolution of the neutrino asymmetry in the
range $1\ MeV\ \stackrel{<}{\sim}\ T\ \stackrel{<}{\sim}\ T_c/2$.
We will discuss this and provide an illustrative
example in section V.

As preparation for the application of the above formalism
to BBN, we need to discuss 
how an asymmetry in $\nu_e$ can be generated
in the context of an overall neutrino mixing scenario. There are
two generic ways of producing a nonzero $L_{\nu_e}$. 
First, $\nu_e-\nu_s$ oscillations can generate $L_{\nu_e}$
directly. Alternatively, $\nu_{\tau} - \nu_s$ (and/or
$\nu_{\mu}-\nu_s$) oscillations
can generate a large $L_{\nu_{\tau}}$ (and/or
$L_{\nu_{\mu}}$) some of which is then transfered to $L_{\nu_e}$
by $\nu_{\tau}-\nu_e$ (and/or $\nu_{\mu}-\nu_e$) oscillations. 

The direct way of generating $L_{\nu_e}$
is only possible in special circumstances.
Either $|\delta m^2_{es}| \gg |\delta m^2_{\tau s}|,
|\delta m^2_{\mu s}|$ or 
$\nu_s$ has significant mixing with $\nu_e$ only.
Only in these circumstances can we consider
$\nu_e - \nu_s$ oscillations in isolation.
For this case we have estimated the effects of the
neutrino asymmetry on BBN by writing a nucleosynthesis
code. We find that $-1.8 \stackrel{<}{\sim} 
\delta N^{eff}_{\nu} \stackrel{<}{\sim} -0.1$
requires a $|\delta m^2_{es}|$ in the
range $0.5 - 7 \ eV^2$.
For $|\delta m^2_{es}| \stackrel{<}{\sim} 0.5\ eV^2$
lepton number is created too late to significantly affect BBN,
while for $|\delta m^2_{es}| \stackrel{>}{\sim}
7 \ eV^2$ lepton number is created so early that it
leads to $\delta N_{\nu} \stackrel{<}{\sim} -1.8$
and thus appears to be too great a modification of BBN
to be consistent with the observations. Note however that 
for $\sin^2 2\theta_0$ large enough, the sterile neutrino can
be excited at temperatures before significant
lepton number is generated (which for $|\delta m^2_{es}| 
\sim 1\ eV$ is $T \stackrel{>}{\sim} 13$ MeV).
This can lead to an increase in the energy density which
can (partially) compensate for a large positive electron 
lepton number.

While the above direct way of generating $L_{\nu_e}$ is
a possibility, we believe that a more interesting possibility
is that $L_{\nu_e}$ is generated indirectly.
As we will show, this mechanism
gives $\delta N_{\nu} \sim - 0.5$ (assuming
$L_{\nu_e} > 0$) for a wide
range of parameters. This mechanism is also the only 
possibility if $|\delta m^2_{\tau s}| \gg
|\delta m^2_{es}|$ or ($|\delta m^2_{\mu s}| \gg
|\delta m^2_{es}|$), assuming that $\nu_s$ mixes with all
three ordinary neutrinos.

\vskip 0.5cm
\noindent
{\bf IV. An example with four neutrinos}
\vskip 0.5cm
Consider the system comprising $\nu_{\tau}, \nu_{\mu}, \nu_e, \nu_s$.
An experimental motivation for the sterile neutrino
comes from the current neutrino anomalies.
There are several ways in which the
sterile neutrino can help solve these problems.
For example, the solar neutrino problem can be solved if
$\delta m^2_{es}/eV^2 \simeq 10^{-6}$ and $\sin^2 2\theta_0 
\simeq 10^{-2}$ (small angle MSW solution)\cite{msw} or if
$10^{-2} \stackrel{<}{\sim} |\delta m^2_{es}|/eV^2 
\stackrel{<}{\sim} 10^{-10}$ and $\sin^2 2\theta_0 \simeq 1$ 
(maximal oscillation solution)\cite{flv}.
Alternatively, $\nu_{\mu} - \nu_s$ oscillations can
solve the atmospheric neutrino problem if $|\delta m^2_{\mu s}|
\simeq 10^{-2}\ eV^2$ and $\sin^2 2\theta_0 \simeq 1$\cite{ana}.

We will assume that
$m_{\nu_{\tau}} \gg m_{\nu_{\mu}}, m_{\nu_e}, m_{\nu_s},$
which means that
\begin{equation}
|\delta M^2| \equiv 
|\delta m^2_{\tau e}| \simeq |\delta m^2_{\tau s}|
\simeq |\delta m^2_{\tau \mu}|
\gg |\delta m^2_{es}|, |\delta m^2_{\mu s}|, |\delta m^2_{\mu e}|.
\end{equation}
With the above assumption, $\nu_{\tau} - \nu_s$ oscillations initially 
create significant $L_{\nu_{\tau}}$ at the temperature 
$T = T_c \simeq 16(|\delta M^2|/eV^2)^{1/6}$.
As before, we will assume that the sign of $L_{\nu_{\tau}}$ is
positive\cite{fn}.
The effect of $\nu_{\tau} - \nu_e$ and $\nu_{\tau} - \nu_{\mu}$ 
oscillations is to generate $L_{\nu_e}$ and $L_{\nu_{\mu}}$
in such a way that 
$L^{(e)} -  L^{(\tau)} = L_{\nu_e} - L_{\nu_{\tau}}
\to 0$ and $L^{(\mu)} - L^{(\tau)} = L_{\nu_{\mu}} - L_{\nu_{\tau}} 
\to 0$, respectively\cite{shi, fv, fn2}. [Note that if $L_{\nu_{\tau}} >
0$, then the MSW resonances for $\nu_{\tau} - \nu_e$ and
$\nu_{\tau} - \nu_{\mu}$ oscillations occur for anti-neutrinos 
(given also our assumption that $m_{\nu_{\tau}} > m_{\nu_{e,\mu}}$)
and so the signs of $L_{\nu_e}$ and $L_{\nu_{\mu}}$ are also positive]. 
However, the rate of change of lepton number due
to collisions, the dominant process at higher $T$, 
is typically too small to efficiently generate
$L_{\nu_e}$ from $L_{\nu_{\tau}}$\cite{shi,fv}.
But, as $L_{\nu_{\tau}}$ becomes large at lower $T$, 
lepton number
can be efficiently transferred by MSW transitions. (When
$L_{\nu_{\tau}} \ll 1$, MSW transitions cannot efficiently
create $L_{\nu_e}$ because $N_{\bar \nu_{\tau}} - 
N_{\bar \nu_{e}} \simeq 0$ and MSW transitions only
interchange $\bar \nu_{\tau}$ with $\bar \nu_e$ without changing their
overall number density).  The rate of change of lepton number due to 
$\bar \nu_{\alpha} - \bar \nu_{\beta}$ oscillations is simply 
given by the difference in rates for which $\bar \nu_{\alpha}$ 
anti-neutrinos and $\bar \nu_{\beta}$ anti-neutrinos pass through 
the resonance (assuming that $\sin^2 2\theta_0 \ll 1$).  
We need to consider the three resonances,
$\bar \nu_{\tau} - \bar \nu_s$, $\bar \nu_{\tau} - \bar \nu_e$ and
$\bar \nu_{\tau} - \bar \nu_{\mu}$, for our system.
We denote the resonance momenta of these resonances
by $p_1$, $p_2$ and $p_3$, respectively. The
rate of change of the lepton numbers due to MSW
transitions is governed approximately by the
differential equations,
\begin{eqnarray}
&{dL_{\nu_{\tau}} \over dT} = - X_1 |{d(p_1/T) \over dT}|
- X_2|{d(p_2/T)\over dT}| 
- X_3|{d(p_3/T)\over dT}|,& \nonumber \\
&{dL_{\nu_{\mu}}\over dT} =  + X_3|{d(p_3/T) \over dT}|,\
{dL_{\nu_{e}}\over dT} =  + X_2|{d(p_2/T) \over dT}|,&\  
\label{day}
\end{eqnarray}
where
\begin{equation}
X_1 \equiv {T\over n_{\gamma}} \left(N_{{\bar \nu}_{\tau}} -
N_{{\bar \nu}_s}\right),\
X_2 = {T \over n_{\gamma}}
\left(N_{{\bar \nu}_{\tau}} - N_{{\bar \nu}_{e}}\right),\ 
X_3 = {T \over n_{\gamma}}
\left(N_{{\bar \nu}_{\tau}} - N_{{\bar \nu}_{\mu}} \right), 
\end{equation}
and the $X_i$ are evaluated at $p = p_i$ $(i = 1,2,3)$.
Note that $X_i$ depends on $T$ through the
ratio $p_i/T$ and through the dependence of the various
chemical potentials on $T$.
Observe that
\begin{equation}
{d(p_i/T)\over dT} = {\partial(p_i/T) \over \partial T} + 
{\partial (p_i/T) \over \partial L_{\nu_{e}}}
{dL_{\nu_{e}}\over dT} +
{\partial (p_i/T) \over \partial L_{\nu_{\mu}}}
{dL_{\nu_{\mu}}\over dT} +
{\partial (p_i/T) \over \partial L_{\nu_{\tau}}}
{dL_{\nu_{\tau}}\over dT}, 
\end{equation}
with
\begin{eqnarray}
& {\partial (p_1/T)\over \partial L_{\nu_{\tau}}} = 
2{\partial (p_1/T)\over \partial L_{\nu_{\mu}}} = 
2{\partial (p_1/T)\over \partial L_{\nu_{e}}} = 
{-2(p_1/T)\over L^{(\tau)}},&  \nonumber \\
& {\partial (p_2/T)\over \partial L_{\nu_{\tau}}} = 
-{\partial (p_2/T)\over \partial L_{\nu_{e}}} = {-(p_2/T)\over
L^{(\tau)} - L^{(e)}},\quad 
{\partial(p_2/T) \over \partial L_{\nu_\mu}} = 0,& \nonumber \\
& {\partial (p_3/T)\over \partial L_{\nu_{\tau}}} = 
-{\partial (p_3/T)\over \partial L_{\nu_{\mu}}} = {-(p_3/T)\over
L^{(\tau)} - L^{(\mu)}},\quad 
{\partial(p_3/T) \over \partial L_{\nu_e}} = 0,& \nonumber \\
&{\partial (p_i/T)\over \partial T} = {-4p_i \over T^2}.&\ 
\end{eqnarray}
By the symmetry of the problem, $L_{\nu_{\mu}} = L_{\nu_e}$,
$p_2 = p_3$ and $dL_{\nu_{\mu}}/dT = dL_{\nu_e}/dT$\cite{happy}. 
Using this simplification, we find that
\begin{equation}
{dL_{\nu_{e}}\over dT} = {dL_{\nu_{\mu}} \over dT}
= {A \over B},
\quad {dL_{\nu_{\tau}}\over dT}=
{\alpha \over y_1} + {\beta \over y_1}{dL_{\nu_e} \over dT},
\label{kl}
\end{equation}
where
$ A = \gamma y_1 + \alpha \delta, \ B = y_1 y_2 - \beta \delta$, with
\begin{eqnarray}
& \alpha = -4X_1\left[ {p_1 \over T^2} \right] - 
8X_2\left[{p_2 \over T^2} \right], \ 
\beta = -2X_1 \left[{p_1 \over TL^{(\tau)}}\right]  
+ 2X_2\left[{p_2 \over T(L^{(\tau)}-L^{(e)})}\right], \nonumber \\
& \gamma = 4X_2\left[{p_2 \over T^2}\right], \
\delta = +X_2\left[{p_2 \over T(L^{(\tau)}-L^{(e)})}\right], \nonumber
\\
& y_1 = 1 + 2X_1\left[{p_1 \over TL^{(\tau)}}\right] 
+ 2X_2\left[{p_2 \over T(L^{(\tau)} - L^{(e)})}\right], \
y_2 = 1 + X_2\left[{p_2 \over T(L^{(\tau)} - L^{(e)})}\right].
\end{eqnarray}
In deriving this equation we have assumed that
$d(p_1/T)/dT < 0$ and
$d(p_2/T)/dT < 0$. Observe that $X_2d(p_2/T)/dT = -A/B$ 
and thus for selfconsistency
Eq.(\ref{kl}) is only valid provided that $A/B < 0$ (given
that $X_2 < 0$).
If $d(p_2/T)/dT > 0$ then Eq.(\ref{kl}) becomes
\begin{equation}
{dL_{\nu_{e}}\over dT} = {dL_{\nu_{\mu}} \over dT}
= {\stackrel{\sim}{A} \over \stackrel{\sim}{B}}
\quad {dL_{\nu_{\tau}}\over dT}=
{\stackrel{\sim}{\alpha} \over \stackrel{\sim}{y_1}} + 
{\stackrel{\sim}{\beta} \over \stackrel{\sim}{y_1}}{dL_{\nu_e} \over dT},
\label{kll}
\end{equation}
where $\stackrel{\sim}{A}, \stackrel{\sim}{B},
\stackrel{\sim}{\alpha}, \stackrel{\sim}{\beta},
\stackrel{\sim}{y_1}$ have the same form as $A, B, \alpha, \beta,
y_1$ except that $X_2 \to - X_2$.  
In this case, $X_2d(p_2/T)/dT = \stackrel{\sim}{A}/\stackrel{\sim}{B}$. 
It follows that Eq.(\ref{kll}) is only selfconsistent provided that
$\stackrel{\sim}{A}/\stackrel{\sim}{B} < 0$.
Observe that $d(p_2/T)/dT$ must be continuous which
means that $d(p_2/T)/dT$ only changes sign when $A$ changes
sign and Eq.(\ref{kl}) maps onto Eq.(\ref{kll}) continuously
because $\stackrel{\sim}{A} = -A$ (and thus $\stackrel{\sim}{A}
= A $ at the point where $A = 0$).
If $d(p_1/T)/dT$ changes sign at some point
$p_1/T = q$ then we must make the
replacement $X_1 \simeq 0$ for $p_1/T < q$ (assuming that
initially $d(p_1/T)/dT < 0$) since the previous MSW transitions
have populated $\nu_s$ for $p_1/T < q$.

In solving Eq.(\ref{kl}), we will assume that 
the initial number of sterile neutrinos
can be neglected (this will be valid for a wide range of
parameters as will be discussed later).  We start 
the evolution of Eq.(\ref{kl}) 
when $T \simeq T_c/2$ (with $T_c$ given by Eq.(\ref{Tc}) for 
$\nu_{\tau} - \nu_s$ oscillations).
There is a range of values of $L_{\nu_{\tau}}$ at this point
which is related to the range of $p_{res}/T$ [Eq.(\ref{range})]
through Eq.(\ref{hird}).
Performing the numerical integration, we find that the final
electron neutrino asymmetry is\cite{sad} 
\begin{eqnarray}
& L^f_{\nu_e}/h \simeq 2.0\times 10^{-2} \ for \ 10 \stackrel{<}{\sim} 
|\delta M^2|/eV^2
\stackrel{<}{\sim} 3000
\nonumber \\
& L^f_{\nu_e}/h \simeq 1.7\times 10^{-2} \ for \ 
|\delta M^2|/eV^2
\stackrel{>}{\sim} 3000
\end{eqnarray}
and recall that $h \equiv T_{\nu_{\alpha}}^3/T_{\gamma}^3$.
We found that $L^f_{\nu_e}$ is approximately independent of 
$p_{res}/T$ for
$p_{res}/T$ in the range given by Eq.(\ref{range}).
We also found numerically that $L^f_{\nu_e}$ is approximately 
independent of the initial value of $L_{\nu_e}$ (at $T \simeq T_c/2$), 
so long as $L_{\nu_e} \stackrel{<}{\sim} L_{\nu_{\tau}}$ at this
temperature (which should be valid since efficient generation of 
$L_{\nu_e}$ does not occur until much lower temperatures where
$L_{\nu_{\tau}}$ has become very large).
In addition, we found that $L^f_{\nu_e}$ is independent 
of the precise value of the initial temperature
(so long as the initial temperature is less than $T_c$ and
$L_{\nu_{\tau}} \ll 1$ at this temperature).
The reason for this independence is simply due to
the fact that significant generation of $L_{\nu_e}$
cannot occur until $L_{\nu_{\tau}}$ becomes large ($\stackrel{>}{\sim}
10^{-2}$). The final asymmetry $L^f_{\nu_e}$ is also 
independent of $\sin^2 2\theta_0$
so long as $\sin^2 2\theta_0 \ll 1$ for aforementioned reasons. 
Finally, and perhaps of most interest, we find that 
$L^f_{\nu_e}$ is almost independent of $|\delta M^2|$ so 
long as $|\delta M^2| \stackrel{>}{\sim} 3 \ eV^2$. 
For $|\delta M^2| \stackrel{<}{\sim} 3 \ eV^2$, $L_{\nu_{\tau}}$ 
does not become large until $T \stackrel{<}{\sim} 1 \ MeV$. 
For temperatures in this range, the effect of $L_{\nu_{\tau}}$ 
cannot be described in terms of chemical potentials because 
the weak interactions are too weak to thermalise the neutrino 
distribution. For this reason, $L^f_{\nu_e}$ should be much 
smaller since the $\bar \nu_{\tau} - \bar \nu_e$ resonance (which 
occurs at a momentum which is always greater than the $\bar 
\nu_{\tau} - \bar \nu_s$ resonance) simply interchanges
almost equal numbers of $\bar \nu_{\tau}$'s and $\bar \nu_e$'s.

We now apply the above analysis to BBN.
Recall that the neutrino oscillations affect
$N^{eff}_{\nu}$ in two ways. 
First, the creation of $L^f_{\nu_e}$
and the related modification of the neutrino momentum distributions
directly affects the nuclear reaction rates which determine the
neutron/proton ratio.  Second, the oscillations can
modify the energy density of the Universe by the excitation of the
sterile neutrino and the modification of the neutrino momentum
distributions due to chemical potentials. 
We first discuss the energy density question.

For $T > T_c$ the $\nu_{\tau} - \nu_s$ oscillations can excite the
sterile neutrino (and anti-neutrino). In Ref.\cite{fv}, a detailed 
study was done
which found that $\rho_{\nu_s}/\rho_{\nu} \stackrel{<}{\sim} 0.6$
provided that 
\begin{equation}
\sin^2 2\theta_0 \stackrel{<}{\sim} 4 \times 10^{-5}\left[
{eV^2 \over |\delta M^2|}\right]^{1/2}.
\end{equation}
Furthermore, $\rho_{\nu_s}/\rho_{\nu} \to 0$ very quickly as
$\sin^2 2\theta_0 \to 0$. In particular, we found that
$\rho_s/\rho_{\nu} \stackrel{<}{\sim} 0.1$
for 
\begin{equation}
\sin^2 2\theta_0 \stackrel{<}{\sim} 5 \times 
10^{-6}\left[ {eV^2 \over |\delta M^2|}\right]^{1/2}.
\label{dd}
\end{equation}
Note that after the lepton number is created, the oscillations
no longer excite significant numbers of sterile neutrinos
until $L_{\nu_{\tau}} \stackrel{>}{\sim} 10^{-2}$.
At this point the $\bar \nu_{\tau} - \bar \nu_s$ oscillations (recall
that we are assuming that $L_{\nu_{\tau}} > 0$)
transfer $\bar \nu_{\tau}  \to 
\bar \nu_s$. The effect of these oscillations on the overall energy
density depends on the temperature 
where $L_{\nu_{\tau}} \stackrel{>}{\sim} 10^{-2}$ occurs, which in
turn depends on $|\delta M^2|$.
There are essentially three regions to consider,
$10 \stackrel{<}{\sim} |\delta M^2|/eV^2 \stackrel{<}{\sim} 3000$,
 $|\delta M^2|/eV^2 \stackrel{<}{\sim} 10$ and
$|\delta M^2|/eV^2 \stackrel{>}{\sim} 3000$.

For $10 \stackrel{<}{\sim} |\delta M^2|/eV^2 \stackrel{<}{\sim} 3000$,
we have numerically calculated the final number and mean energies of
$\bar \nu_{\tau}, \bar \nu_s, \bar \nu_e, \bar \nu_{\mu}$ (the
number and energy densities of the neutrinos are approximately 
unchanged in this region). Normalizing the number density to
the number of neutrinos when $\mu_{\nu} = 0$, 
$n_0 \equiv {3 \over 4}\zeta (3)T^3/\pi^2$, we find
\begin{equation}
{n_{\bar \nu_e} \over n_0} = 
{n_{\bar \nu_{\mu}} \over n_0} \simeq 0.95,\quad
{n_{\bar \nu_{\tau}} \over n_0} \simeq 0.44,\quad
{n_{\bar \nu_{s}} \over n_0} \simeq 0.66.
\end{equation}
Note that the total number is approximately unchanged (i.e.
$0.95\times 2 + 0.44 + 0.66 \simeq 3)$. We find the final mean 
energy for the $\bar \nu_s$, $\langle E_s \rangle$, to be 
slightly less than 
than the mean energy for a Fermi-Dirac distribution 
with $\mu_{\nu} = 0$,
$\langle E_{FD} \rangle \simeq 3.15T$ 
[$\langle E_s \rangle/\langle E_{FD} \rangle 
\simeq 0.88$].  For this reason there is a small overall change 
in energy density, equivalent to about $\delta N^{eff}_{\nu} \simeq 
-0.05$.  For $|\delta M^2| \stackrel{>}{\sim} 3000 \ eV^2$, 
$L^f_{\nu_{\tau}}$ is reached for $T \stackrel{>}{\sim} 
T_{dec}^{\tau}$ and so
$\mu_{\nu_{\tau}} \simeq - \mu_{\bar \nu_{\tau}}$.
In this case, there is an additional contribution to the
energy density coming from the $\nu_{\tau}$ neutrinos
due to the negative chemical potential $\mu_{\nu_{\tau}}$.
In this case we find that the overall change in the energy
density is considerably larger and equivalent to 
$\delta N^{eff}_{\nu} \simeq 0.4$.
Finally, for $|\delta M^2| \stackrel{<}{\sim} 10 \ eV^2$,
the change in the energy density quickly becomes completely
negligible because the weak interactions are unable to 
thermalise the neutrino distributions. The oscillations
simply transfer $\bar \nu_{\tau} $ to $\bar \nu_s$ and the
total number and energy density remain unchanged.

We now turn to the effect of $L_{\nu_e}^f$ and the corresponding 
modification of the momentum distributions on $N^{eff}_{\nu}$
through nuclear reaction rates.
For $10 \stackrel{<}{\sim} |\delta M^2|/eV^2
\stackrel{<}{\sim} 3000$, 
the distribution of $L^f_{\nu_e}$ can
be approximately described by chemical potentials
$\stackrel{\sim}{\mu}_{\bar \nu} \ \simeq 0.06$ and
$\stackrel{\sim}{\mu}_{\nu} \ \simeq 0$. For
$|\delta M^2|\stackrel{>}{\sim} 3000 \ eV^2$, the
lepton number is created above the chemical decoupling 
temperature. In this case the distribution of $L^f_{\nu_e}$ can
be approximately described by chemical potentials
$\stackrel{\sim}{\mu}_{\bar \nu} \ \simeq 0.025$ and
$\stackrel{\sim}{\mu}_{\nu} \ \simeq -0.025$. 
We find that for $|\delta M^2| \stackrel{>}{\sim} 
10 \ eV^2$, $L = L^f_{\nu_e}$ is reached for $T \stackrel{>}{\sim}
1.5\ MeV$. Thus to a good approximation the chemical potentials
$\stackrel{\sim}{\mu}_{\nu_e}, \stackrel{\sim}{\mu}_{\bar \nu_e}$
are approximately constant during the nucleosynthesis era.
Using our BBN code, we find that the modification of $Y_P$ due
to the chemical potentials is 
$\delta Y_P \simeq -0.005$ for 
$\stackrel{\sim}{\mu}_{\bar \nu}\ \simeq 0.06,\
\stackrel{\sim}{\mu}_{\nu}\ \simeq 0$ 
and $\delta Y_P \simeq -0.006$ for
$\stackrel{\sim}{\mu}_{\bar \nu}\ \simeq 0.025, \
\stackrel{\sim}{\mu}_{\nu}\ \simeq -0.025$. From 
Eq.(\ref{2}), this translates into a 
reduction of the effective number of neutrino degrees of freedom
during nucleosynthesis. Including the effects of the 
change in energy density discussed earlier, we find that
\begin{eqnarray}
\delta N^{eff}_{\nu} \simeq - 0.5 \ for \ 10 \stackrel{<}{\sim} 
|\delta M^2|/eV^2 \stackrel{<}{\sim} 3000 \nonumber \\
\delta N^{eff}_{\nu} \simeq - 0.1 \ for \  
|\delta M^2|/eV^2 \stackrel{>}{\sim} 3000.
\label{res}
\end{eqnarray}
For this result,
we have considered the case of negligible excitation of 
sterile neutrinos for temperatures above $T_c$, that is,
Eq.(\ref{dd}) has been assumed.
Note that if we had assumed that $L_{\nu_{\tau}}$ was
negative instead of positive, then the sign of $L_{\nu_e}$ is 
also negative and the change in $Y_P$ due to the asymmetry
is opposite in sign as well.
This leads to $\delta N^{eff}_{\nu} \simeq +0.4 \ (0.9)$ for
$10 \stackrel{<}{\sim} 
|\delta M^2|/eV^2 \stackrel{<}{\sim} 3000$
($|\delta M^2|/eV^2 \stackrel{>}{\sim} 3000$). 

In our analysis we have neglected the effects of the
$\nu_{\mu} - \nu_s$, $\nu_{\mu} - \nu_e$ and $\nu_e - \nu_s$
oscillations. It is usually possible to neglect these oscillations
if $|\delta m^2| \ll |\delta M^2|$
because the lepton number created by $\nu_{\tau} - \nu_s$ oscillations
is large enough to suppress the oscillations which have
much smaller $\delta m^2$.  Of course, in some circumstances these 
oscillations cannot be neglected. For example, in Ref.\cite{fv},
we showed that the effects of maximal $\nu_{\mu} - \nu_s$ oscillations
with $|\delta m^2_{\mu s}| \simeq 10^{-2}\ eV^2$ (as suggested
by the atmospheric neutrino anomaly\cite{ana})
can only be neglected if $|\delta m^2_{\tau s}| \stackrel{>}{\sim}
30 \ eV^2$.  Interestingly, this parameter space overlaps 
considerably with the parameter space where $\delta N^{eff}_{\nu} 
\simeq -0.5$, according to Eq.(\ref{res}).
Note that this parameter space is also suggested if
the tau neutrino is a significant component of dark matter.

\vskip 2.5cm
\noindent
{\bf V. Evolution of lepton number from the
exact quantum kinetic equations}
\vskip 0.5cm
In this section we study the evolution of
the neutrino asymmetry by numerically integrating the
exact quantum kinetic equations\cite{bm}.
This formalism allows a near exact calculation 
to be performed which is valid at both high and low
temperatures.
As we have discussed, for high temperatures $T \stackrel{>}{\sim}
T_c$ the evolution of lepton number is dominated by collisions
(assuming $|\delta m^2| \stackrel{>}{\sim} 10^{-4} \ eV^2$)
while at lower temperatures, the evolution of lepton number
is dominated by oscillations between collisions (MSW effect).

The system of an active neutrino oscillating with a sterile
neutrino can be described by a density
matrix\cite{bm, hist}.  Below we very briefly outline this formalism.
The density matrices describing an ordinary neutrino
of momentum $p$ oscillating with a sterile neutrino 
are given by
\begin{equation}
\rho_{\nu}(p) = {1 \over 2}P_0(p)[1 + {\bf P}(p) \cdot \sigma],
\  \rho_{\bar \nu}(p) = {1 \over 2}\bar P_0(p) 
[1 + {\bf \bar P}(p) \cdot \sigma ],
\label{4}
\end{equation}
where ${\bf P}(p) = P_x(p) \hat{x} + P_y(p) \hat{y} + P_z(p) \hat{z}$.
(It will be understood throughout this section that the density
matrices and the quantities $P_i$ also depend on time $t$ or,
equivalently, temperature $T$.)
The number distributions of $\nu_{\alpha}$ and $\nu_s$ are given by
\begin{equation}
N_{\nu_{\alpha}} = {1 \over 2}P_0(p)[1 + P_z(p)]N_{\nu_{\alpha}}^{eq},\
N_{\nu_s} = {1 \over 2}P_0(p)[1 - P_z(p)]N_{\nu_{\alpha}}^{eq}
\label{5}
\end{equation}
where 
\begin{equation}
N_{\nu_{\alpha}}^{eq} = {1 \over 2\pi^2}{p^2 \over 
1 + exp\left({p+\mu_{\nu}\over T}\right)},
\end{equation}
is the equilibrium number distribution. Note that
there are analogous equations for the anti-neutrinos (with
${\bf P}(p) \to {\bf \bar P}(p)$ and $P_0 \to \bar P_0$). 
The evolution of $P_0(p)$ and ${\bf P}(p)$ are governed by the equations
\cite{bm},
\begin{eqnarray}
{\partial \over \partial t}{\bf P}(p) & = & {\bf V}(p) \times {\bf P}(p) + 
[1 - P_z(p)]\left[{\partial \over \partial t}\ln P_0(p)\right] {\bf \hat{z}}
\nonumber \\
&-& [D(p) + {d \over dt}\ln P_0(p)][P_x(p) {\bf \hat{x}} + 
P_y(p) {\bf\hat{y}}], 
\nonumber \\
{\partial \over \partial t}P_0(p) & \simeq & R(p).
\label{6}
\end{eqnarray}
The quantity ${\bf V}(p)$ is given by
\begin{equation}
{\bf V}(p) = \beta(p) {\bf \hat{x}} + \lambda(p) {\bf \hat{z}},
\label{7}
\end{equation}
where $\beta(p)$ and $\lambda(p)$ are defined by
\begin{equation}
\beta(p) = {\delta m^2 \over 2p} \sin2\theta_0, \
\lambda(p) = -{\delta m^2 \over 2p }[\cos2\theta_0 - b(p) \pm a(p)],
\label{8}
\end{equation}
in which the $+(-)$ sign corresponds to neutrino (anti-neutrino) 
oscillations.  The dimensionless variables $a(p)$ and $b(p)$ contain
the matter effects and are given in Eq.(\ref{ab}).
The quantity $D(p)$ is the quantum damping 
parameter resulting from the collisions of the neutrino with the
background.  According to Ref.\cite{stod}, 
the damping parameter is half of the total collision frequency,
i.e. $D(p) = \Gamma_{\nu_{\alpha}}(p)/2$.
Finally, note that in Eq.(\ref{6}) the function $R(p)$ is 
related to $\Gamma_{\nu_{\alpha}}(p)$ and its specific definition
is given in Ref.\cite{bm}.
For temperatures above $1$ MeV, we can make the useful approximation
of setting $N_{\nu_{\alpha}} = N_{\nu_{\alpha}}^{eq}$
and $N_{\bar \nu_{\alpha}} = N_{\bar \nu_{\alpha}}^{eq}$. This 
means that $P_0(p) = 2/[1 + P_z(p)], \ \bar P_0(p) = 
2/[1 + \bar P_z(p)]$ and consequently 
\begin{equation}
{\partial P_0(p) \over \partial t} = 
{-2 \over [1 + P_z(p)]^2}{\partial P_z(p) \over \partial t},\ 
{\partial \bar P_0(p) \over \partial t} = {-2 \over [1 + \bar P_z(p)]^2}
{\partial \bar P_z(p) \over \partial t}.
\label{late}
\end{equation}

For the numerical work, the continuous variable
$p/T$ is replaced by a finite set of momenta $x_n \equiv p_n/T$
(with $n = 1,...,N$). 
The variables ${\bf P}(p)$ and $P_0(p)$ are replaced
by the set of $N$ variables ${\bf P}(x_n)$ and $P_0(x_n)$. The
evolution of each of these variables is governed by Eqs.(\ref{6}),
where for each value of $n$, the variables ${\bf V}(p)$ and $D(p)$ 
are replaced by ${\bf V}(x_n)$ and $D(x_n)$. 
Thus, the oscillations of the neutrinos and anti-neutrinos can be
described by $8N$ simultaneous differential equations.

The rate of change of lepton number is given by
\begin{equation}
{dL_{\nu_{\alpha}} \over dt} = {d \over dt}\left[{(n_{\nu_{\alpha}} 
- n_{\bar \nu_{\alpha}})\over n_{\gamma}} \right] = - {d \over dt}
\left[ {(n_{\nu_s} - n_{\bar \nu_s}) \over n_{\gamma}}\right].
\end{equation}
Thus, using Eq.(\ref{5}),
\begin{equation}
{dL_{\nu_{\alpha}} \over dt} = 
{1 \over 2}{d \over dt}\left[{1 \over n_{\gamma}}\int 
\left[\bar P_0(1 - \bar P_z)N_{\bar \nu_{\alpha}}^{eq} 
- P_0(1 - P_z)N_{\nu_{\alpha}}^{eq} \right]
dp \right].
\end{equation}
Taking the time differentiation inside the integral we find that
\begin{equation}
{dL_{\nu_{\alpha}} \over dt} \simeq 
{1 \over 2}\int \left(
{\partial [\bar P_0(1 - \bar P_z)] \over \partial t}
{N_{\bar \nu_{\alpha}}^{eq}\over
n_{\gamma}}
- {\partial [P_0(1 - P_z)] \over \partial t}{N_{\nu_{\alpha}}^{eq} \over
n_{\gamma}} \right) dp
\label{er}
\end{equation}
where we have used the 
result that $N_{\nu_{\alpha}}^{eq}dp/n_{\gamma}$ is
approximately independent of $t$.
Expanding this equation using Eq.(\ref{late}), we find
\begin{equation}
{dL_{\nu_{\alpha}} \over dt} =
{1 \over n_{\gamma}}\int \left(
N^{eq}_{\nu}{2 \over [1 + P_z]^2}
{\partial P_z \over \partial t}
- N^{eq}_{\bar \nu}{2 \over [1 + \bar P_z]^2}
{\partial \bar P_z \over \partial t}
\right)dp
\label{erer}
\end{equation}

Equations \ref{6} and \ref{erer} can be numerically integrated to
obtain the evolution of $L_{\nu_{\alpha}}$\cite{ffn, xx}. 
We illustrate this with an example. For
definitness we will consider the $\nu_{\tau}, \nu_s$ system.
In Figure 1, we take $\delta m^2_{\tau s} = -10$ and $\sin^2 2\theta_0
= 10^{-9}$ (we set $\eta = 4 \times 10^{-10}$ and
took $L_{\nu_{\alpha}} = 0$ initially\cite{lad}).
The result of numerically integrating Eq.(\ref{6}) and Eq.(\ref{erer})
is shown in the figure by the dashed-dotted line.
Also shown in Figure 1 (dashed line) is the ``static 
approximation'' (Eqs.(94) and (93) of Ref.\cite{fv}). 
As discussed in Ref.\cite{fv}, the static approximation assumes
that the system is sufficiently smooth and that the dominant
contribution to the rate of change of lepton number is collisions.
As shown in Figure 1, the static approximation is a good approximation
at high temperatures.  However, as discussed in Ref.\cite{fv},
the static approximation does not include the MSW effect which
is the dominant physical process at low temperatures.
As expected the MSW effect keeps $L_{\nu_{\tau}}$ growing like 
$L_{\nu_{\tau}} \sim 1/T^4$ for much lower temperatures.
We have also checked our simplified Equation (\ref{cat2}) for
the evolution of lepton number due to MSW transitions.
We started the evolution of this equation at $T = T_c/2 \simeq
13.5\ MeV$ with the value of $L_{\nu_{\tau}}$ at this point 
obtained from the quantum kinetic equations of $L_{\nu_{\tau}}
\simeq 2.92\times 10^{-5}$.  The subsequent evolution of $L_{\nu_{\tau}}$
obtained from numerically integrating Eq.(\ref{cat2})
is given in Figure 1 by the solid line.
As the figure shows, Eq.(\ref{cat2}) is a very good
approximation for the evolution of the neutrino
asymmetry at low temperatures.
This provides a useful check 
of the validity of the approximate approach used in section IV
for the $\nu_e, \nu_{\mu}, \nu_{\tau}, \nu_s$ four flavour
system. 

It is instructive to examine the evolution of the neutrino
and anti-neutrino resonance momenta.
Recall that the resonance for neutrinos
occurs when $b - a = \cos2\theta_0$ while
the resonance for anti-neutrinos 
occurs when $b + a = \cos2\theta_0$.
Let us write 
\begin{equation}
b = \lambda_1 p^2,\ a = \lambda_2 p,
\end{equation}
where $\lambda_1$ and $\lambda_2$ are independent
of $p$ and can be obtained from Eq.(\ref{ab}).
Note that $\lambda_1, \lambda_2 > 0$ given
that $\delta m^2_{\alpha s} < 0$ and 
assuming $L^{(\alpha)} > 0$.
Solving the resonance conditions $b \pm a = 
\cos2\theta_0$, we find that
the resonance momenta satisfy 
\begin{eqnarray}
& p_{res} = {\lambda_2 + \sqrt{\lambda^2_2 + 4\lambda_1 \cos2\theta_0}
\over 2\lambda_1} \ for \ neutrinos, \nonumber \\
& p_{res} = {-\lambda_2 + \sqrt{\lambda^2_2 + 4\lambda_1 \cos2\theta_0}
\over 2\lambda_1}
\ for \ anti-neutrinos.
\end{eqnarray}
In Figure 2, we have plotted the evolution of the resonance 
momenta for the neutrinos and anti-neutrinos.
As this example illustrates, the neutrino resonance momentum moves
to very high values as $T \stackrel{<}{\sim} T_c$ while
the anti-neutrino resonance momentum moves to very
low values (which in this example is $p_{res}/T \simeq 0.6$ for
$T \simeq T_c/2$).
We have found that this behaviour is quite general, with 
the anti-neutrino resonance $p_{res}/T$ in the range Eq.(\ref{range}) 
at $T = T_c/2$ as $\sin^2 2\theta_0$ and $\delta m^2$ are varied. 

\vskip 0.5cm
\noindent
{\bf VI Conclusion}
\vskip 0.5cm

In summary, we have extended previous work on neutrino
oscillation generated lepton number in the early Universe
by studying the evolution of lepton number at low
temperatures where the MSW effect is important.
We applied this work to examine the implications of
the neutrino asymmetry for BBN in two illustrative
models. In the first model, electron neutrino asymmetry
was created directly by $\nu_e - \nu_s$ oscillations
while in the second model the electron neutrino
asymmetry was created indirectly by the reprocessing of
a tau neutrino asymmetry.

One result of this study is that the naive conclusion that
sterile neutrinos only increase the effective number
of neutrino species ($N^{eff}_{\nu}$) during the nucleosynthesis era is
actually wrong. Neutrino asymmetries generated by neutrino 
oscillations can naturally lead to a decrease in $N^{eff}_{\nu}$.
Furthermore in the case where the electron neutrino asymmetry is transferred
from the tau or mu neutrino asymmetries, the electron neutrino
asymmetry is approximately independent of $|\delta m^2|$ and 
$\sin^2 2\theta_0$ for a wide range of parameters.
This leads to a prediction of $\delta N_{\nu}^{eff} \simeq -0.5$
if the asymmetry is positive
for an interesting class of models. Remarkably, this
prediction is supported by some recent observations
which actually suggest $N^{eff}_{\nu} < 3$.

\vskip 0.8cm
\noindent
{\bf Acknowledgements}
\vskip 0.4cm
\noindent
R.F. is an Australian Research Fellow. R.R.V. is supported
by the Australian Research Council.

\newpage
\noindent
{\large \bf Figure Captions}
\vskip 0.5cm
\noindent
Figure 1.  The evolution of the $\nu_{\tau} - \nu_s$ 
oscillation generated lepton number
asymmetry, $L_{\nu_{\tau}}$.
We have taken by way of example, the 
parameter choice $\delta m^2 = -10\ eV^2$ and
$\sin^2 2\theta_0 = 10^{-9}$.
The dashed-dotted line is the result of the numerical
integration of the quantum kinetic equations [Eq.(\ref{erer})
and Eq.(\ref{6})].  The solid line is the result from the 
numerical integration
of Eq.(\ref{cat2}) while the dashed line
is the static approximation developed in Ref.\cite{fv}.
\vskip 0.5cm
\noindent
Figure 2.  
The evolution of the neutrino (dashed line) and 
anti-neutrino (solid line) resonance
momenta for the example of Figure 1. 

\begin{thebibliography}{99}
\bibitem{notation}
In Ref.\cite{fv} we denoted $a \equiv a(p,T)$ by $a^p$, and
we reserved the symbol $a$ for the momentum averaged
quantity $\langle a \rangle$ in which the momentum $p$ is
replaced by its thermal average $\langle p \rangle \simeq 3.15 T$.
Similar remarks pertain to $b$.
We have also used a slightly different definition for the
parameters $A_{\alpha}$.
\bibitem{fn}
The sign of the asymmetry $L_{\nu_{\alpha}}$ generated by
$\nu_{\alpha} - \nu_s$ oscillations depends on the sign of
the initial asymmetries $\eta$ and $L_{\nu_{\beta}}$ (with
$\beta \neq \alpha$)\cite{fv}. Assuming that $m_{\nu_{\tau}} \gg 
m_{\nu_{\mu}}, m_{\nu_e}, m_{\nu_s}$, then
$sign (L_{\nu_{\alpha}}) = sign(\stackrel{\sim}{\eta})$
where $\stackrel{\sim}{\eta}\ \equiv \eta + L_{\nu_e} +
L_{\nu_{\mu}} + L_{\nu_{\tau}} - L_{\nu_{\alpha}}$\cite{fv}.
Thus, for example, $L_{\nu_{\alpha}}$ should be positive if 
the initial $L_{\nu_e}, L_{\nu_{\mu}}, L_{\nu_{\tau}} > 0$.
\bibitem{binh}
This is because collisions also keep
$p_{res}/T$ approximately constant and in the range Eq.(\ref{range})
(for $T < T_c$) except at very low temperatures 
where the collision rate is very low.
\bibitem{fn4}
Note that in the case $\alpha = e$, the laboratory bound
$m_{\nu_e} \stackrel{<}{\sim} 3\ eV$ (assuming a Majorana
mass) implies that
$|\delta m^2_{es}| \stackrel{<}{\sim} 9\ eV^2$. 
Also note that the cosmology energy bound suggests that
$m_{\nu_{\mu}}, m_{\nu_{\tau}} \stackrel{<}{\sim}
40 \ eV$, which implies $|\delta m^2_{\mu s}|, |\delta m^2_{\tau s}|
\stackrel{<}{\sim} 1600 \ eV^2$.
\bibitem{fn2}
The formalism for ordinary - ordinary, $\nu_{\alpha} - \nu_{\beta}$, 
neutrino oscillations is similar to ordinary - sterile 
neutrino oscillations. For ordinary - ordinary  neutrino oscillations
we only need to replace $V_{\alpha}$ by $V_{\alpha} - V_{\beta}$
(actually there are also other effects, but like
the $b$ term of the potential $V$ they are order $G_F^2$ effects,
and they can approximately be neglected in our analysis).
The effect of replacing $V_{\alpha}$ by $V_{\alpha}-V_{\beta}$
can be taken into account by simply replacing $L^{(\alpha)}$
in $a$ by $L^{(\alpha)} - L^{(\beta)}$.
\bibitem{happy}
Actually if $p_2$ is exactly equal to $p_3$ then the two
component formalism will not be strictly valid. However,
$p_2$ will be only approximately $p_3$ because,
for example, $\delta m^2_{\tau \mu}$ will be slightly
different to $\delta m^2_{\tau e}$ due to the
$m_{\nu_{\mu}}, m_{\nu_e}$ mass difference.
\bibitem{sad}
Note that 
we found numerically that for 
$|\delta M^2| \stackrel{<}{\sim} 350\ eV^2$, the evolution 
of $A$ and $B$ is such that $A < 0$, $B <0$ (with 
$d(p_2/T)/dT < 0$) occurs at some temperature during the evolution.
At this point there is no solution for $dL_{\nu_e}/dT$ since
Eq.(\ref{kl}) cannot be applied selfconsistently.  For points 
in this region,
$L_{\nu_e}$ is increasing so rapidly that our simplified
Equations (\ref{day}) are not valid. For
points in this region, we obtained $dL_{\nu_e}/dT$ from
Eq.(\ref{day}) but with $d(p_2/T)/dT$ evaluated at the previous
integration step.
\bibitem{ffn}
In Figure 1 we have plotted $|L_{\nu_{\tau}}|$.
We found that $L_{\nu_{\tau}}$ changes sign at the temperature
$T = T_c$.  As we discussed in Ref.\cite{fv}, this behaviour
is expected.
\bibitem{xx}
In integrating Eq.(\ref{6}) and Eq.(\ref{erer}) we have
only evaluated these equations for momenta in the region 
around the neutrino and anti-neutrino resonances.
\bibitem{lad}
Note that the evolution of $L_{\nu_{\alpha}}$ is essentially
independent of the initial conditions so long as $L_{\nu_{\alpha}}
\stackrel{<}{\sim} 10^{-5}$\cite{fv}.
\end{thebibliography}
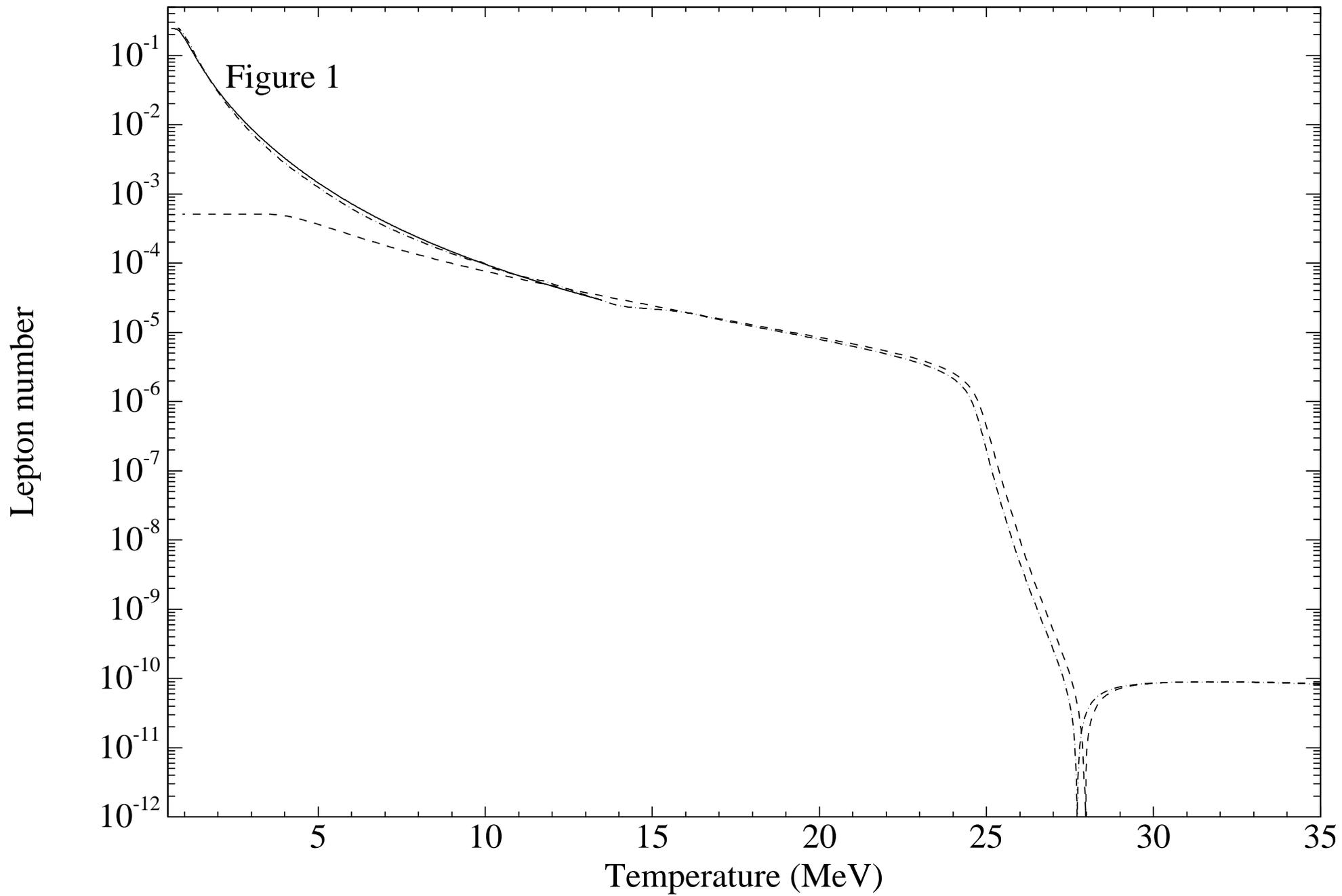
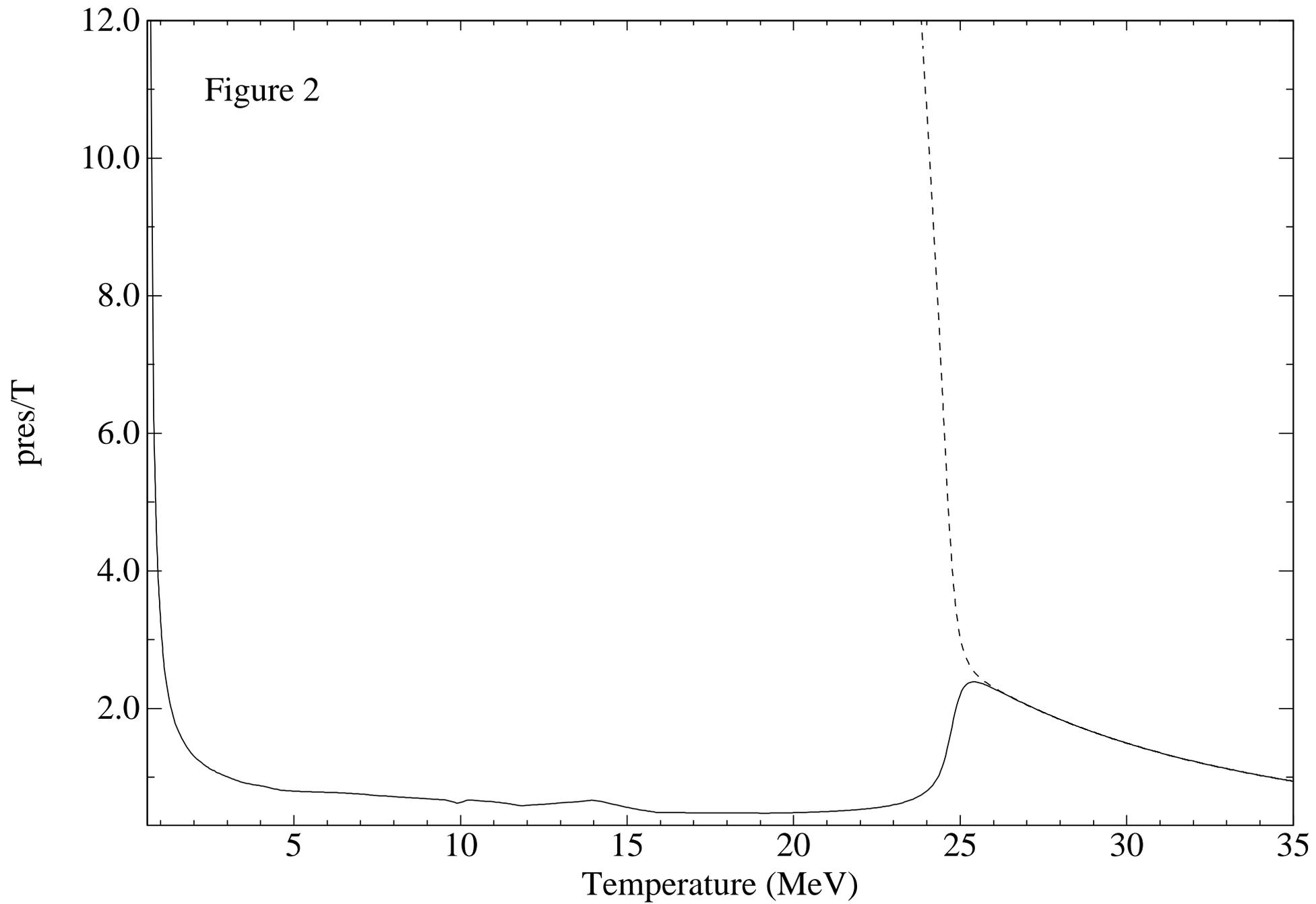
\end{document}